%% file: KDP_Pressure_arxiv.tex
\documentclass[prl,showpacs,floatfix,twocolumn,byrevtex,longbibliography,superscriptaddress]{revtex4-2}
\usepackage{amsmath}
\usepackage{amssymb}
\usepackage{amstext}
\usepackage{amsopn}
\usepackage{amsfonts}
\usepackage{amsxtra}
\usepackage[english]{babel}
\usepackage{graphicx}
\usepackage{float}
\usepackage{siunitx}

\usepackage{xcolor}
\usepackage[version=3]{mhchem} 

\newcommand{\kdp}{\ce{KH2PO4}}
\newcommand{\icm}[1]{\SI{#1}{\centi\meter^{-1}}}
\newcommand{\K}[1]{\SI{#1}{\kelvin}}

\newcommand{\GPa}[1]{\SI{#1}{\giga\pascal}}
\newcommand{\microns}[1]{\SI{#1}{\micro\meter}}

\begin{document}

\title{Revealing the \kdp\ soft-mode coupling mechanism with infrared spectroscopy under pressure}

\author{D. Santos-Cottin}
\affiliation{Department of Physics, University of Fribourg, 1700 Fribourg, Switzerland}

\author{S. Nasrallah}
\affiliation{Department of Physics, University of Fribourg, 1700 Fribourg, Switzerland}

\author{F. Capitani}
\affiliation{Synchrotron Soleil, L'Orme des Merisiers - Départementale 128 - F-91190 Saint-Aubin, France}

\author{P. Simon}
\affiliation{CNRS, CEMHTI UPR3079, Univ. Orléans, F-45071 Orléans, France}


\author{C. C. Homes}
\affiliation{National Synchrotron Light Source II, Brookhaven National Laboratory, Upton, New York 11973, USA}

\author{A. Akrap}
\affiliation{Department of Physics, University of Fribourg, 1700 Fribourg, Switzerland}

\author{R. P. S. M. Lobo}
\email[]{lobo@espci.fr} 
\affiliation{LPEM; ESPCI Paris, PSL Research University; CNRS; Sorbonne Universit\'e, F-75005 Paris, France}

\date{\today}
\begin{abstract}
We measured the far-infrared reflectivity of a \kdp\ single crystal up to pressures of \GPa{2} in the ferroelectric and paraelectric phases.
We find that the $\nu_4$ vibrational mode of the \ce{PO4} tetrahedron is strongly affected by the applied pressure.
At ambient pressure this phonon is destabilized by the presence of the H ions and hence shows a highly damped character, beyond the phonon propagation threshold.
Applying a pressure close to \GPa{0.6} makes this phonon clearly underdamped. 
Its behavior closely follows the soft-mode behavior observed in Raman spectroscopy.
Our results solve a long standing open problem, demonstrating that the $\nu_4$ mode is the excitation mediating the coupling of the hydrogen network to the lattice modes that create the ferroelectic polarization in \kdp.
\end{abstract}
\maketitle

%
%

The riches of ferroelectrics and new phenomena arising from their large dielectrict constant are closely tied to the soft mode mechanism in these materials \cite{Kamba2021}.
Soft-mode-driven remarkable properties have recently brought long-known materials with lattice instabilities into the limelight.
The soft-mode and the proximity to the ferroelectric instability are the key mechanism for the extreme dilute superconductivity in \ce{SrTiO3} \cite{Lin2014,Rischau2022}.
\ce{KTaO3} is an incipient ferroelectric \cite{Shirane1967,Zelezny2004,Ichikawa2005} where the soft-mode underpins new opportunities in spintronics \cite{Gupta2022} and 2D superconductivity \cite{Liu2023}. 
In this paper we study a classic ferroelectric and find a missing link in its peculiar soft-mode.

\kdp\ (KDP) is a ferroelectric material with a Curie temperature of $T_{\rm C} \simeq \K{123}$ \cite{Busch1938}.   
It has a tetragonal ($I\overline{4}2d$) paraelectric phase, which becomes orthorhombic ($Fdd2$) in the ferroelectric state. 
Each H atom in KDP connects two O atoms. 
\citeauthor{Slater1941} \cite{Slater1941} proposed that the proton sitting in the middle of the O-H-O bonding would be an average position. 
In his model the H atom stays closer to one of the O atoms in the \ce{PO4} tetrahedron. 
He found that 6, out of the  16 possible configurations, are energetically more stable. 
Looking at the free energy of the system he calculated an electrical polarization that remains constant below a critical temperature $T_{\rm C}$. 
\citeauthor{Slater1941}'s order-disorder theory gives an overall good description of KDP but fails in two important points. 
The first problem is the constant, temperature independent, polarization below $T_{\rm C}$. 
The second is the lack of an explanation for why $T_{\rm C}$ increases to \K{222} when deuterium substitutes for hydrogen. 

\citeauthor{Pirenne1949} \cite{Pirenne1949} incorporated the kinetic energy of the proton vibrations into \citeauthor{Slater1941}'s picture, thus explaining the difference in $T_{\rm C}$ between H and D materials. 
Elaborating on \citeauthor{Pirenne1949}'s ideas, \citeauthor{Blinc1960} \cite{Blinc1960} placed the proton in a double well potential and proposed that the proton could tunnel between the two wells. 
He assumed a symmetric potential in the high temperature phase and asymmetric wells in the ferroelectric phase. 
The tunneling probability of D being much smaller than that for H, $T_{\rm C}$ is higher for the former material. 
Within this formalism, KDP is no longer a simple order-disorder ferroelectric but it rather has a mixed character with a soft-mode composed of a proton collective mode strongly coupled to an optical phonon and to an acoustic mode.

\begin{figure}
	\includegraphics[width=8cm]{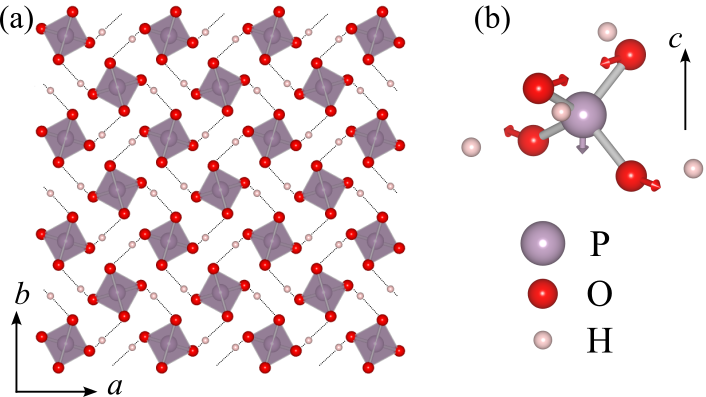}
  	\caption{(color online) (a) KDP structure, with K atoms suppressed, showing the proton position with respect to \ce{PO4} tetrahedra in the orthorhombic ferroelectric phase. 
    (b) Atomic displacements for the $\nu_4$ vibration mode of the \ce{PO4} tetrahedron.}
  	\label{figStructure}
\end{figure}
Curiously, according to this description, ferroelectricity in KDP comes from a  Hamiltonian dominated by the proton network with heavy ions entering as a perturbative term. 
Nevertheless, the ferroelectric polarization comes from the displacement of heavier ions. 
Figure~\ref{figStructure}~(a) shows the proton network in KDP and its connections with the \ce{PO4} tetrahedron in the ferroelectric phase. 
Remarkably, the ferroelectric polarization forms along the $c$-axis --- a direction perpendicular to the $ab$-plane, which contains the  proton tunneling mode.

Gradually, \citeauthor{Blinc1960}'s theory settled as the accepted model for the ferroelectric transition in KDP. 
Some experimental evidence supporting this picture started to emerge with Raman measurements of an overdamped soft mode \cite{Kaminow1968}. 
Later, neutron structural studies showed the existence of the proton double-well potential \cite{Nelmes1984}. 
More recently, neutron Compton scattering \cite{Reiter2002} measured the momentum distribution of hydrogen in \kdp. This data, in conjuction with ab-initio calculations \cite{Zhang2001} highlight the importance of the coupling between the proton tunneling and the \ce{PO4} tetrahedron \cite{Sugimoto1991,Bussmann-Holder1998}.

The importance of the H coupling to the \ce{PO4} group had already been put forward by \citeauthor{Simon1985} \cite{Simon1985,Simon1988}. 
They showed that the infrared reflectivity of \ce{RbH2PO4} has an overdamped soft mode, seen by Raman, and that the same mechanism occurs in \ce{KH2PO4} and \ce{KH2AsO4}. 
Their spectra also shows a highly-damped $\nu_4$ mode of the \ce{PO4} tetrahedra. 
Figure~\ref{figStructure}~(b) shows the atomic displacements of this mode. 
They proposed that the nearby proton tunneling between their two equilibrium positions destabilizes the $\nu_4$ mode, which softens its frequency and increases its damped character. 
As this vibration has a dipole moment, they conjectured that this mode would be the missing link coupling the proton network to the external translation overdamped mode. 

Here, we utilized infrared measurements under pressure to give a definite proof that this phonon is indeed coupled to the soft mode. 
\citeauthor{Peercy1973} showed that the overdamped soft mode measured by Raman scattering becomes underdamped under an applied pressure close to \GPa{1} \cite{Peercy1973}. 
We find that the damping of the $\nu_4$ mode also strongly decreases under similar pressures in both the ferroelectric and paraelectric phases. 
This shows that the two modes are coupled and that the $\nu_4$ vibration of the \ce{PO4} group is the excitation responsible for coupling the hydrogen network to the soft mode of \kdp.

%
%

%
%
\begin{figure}
	\includegraphics[width=8cm]{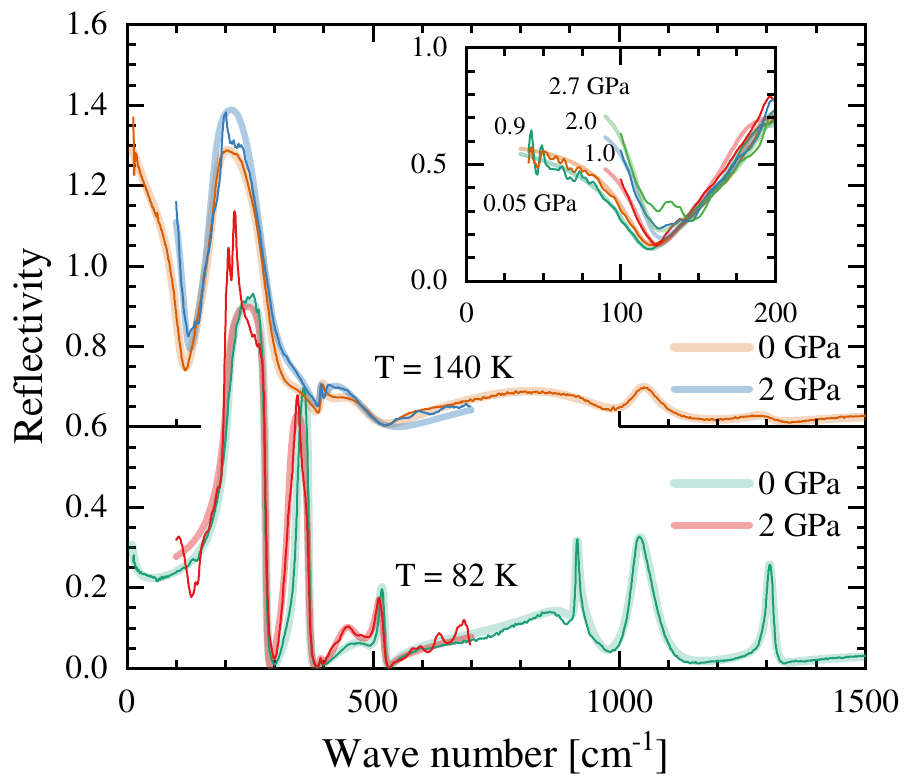}
  	\caption{(color online) Infrared reflectivity polarized along the ferroelectric $c$-axis direction in the paraelectric (\K{140}) and ferroelectric (\K{82}) phases. 
    The ambient pressure data were recorded in the full shown spectral range, whereas data at \GPa{2} were taken between 100 and \icm{600}. 
    The thicker, light-colored lines are fits taking a dielectric function of the form given by Eq.~(\ref{eqLST}) and the thinner lines are the measurements. 
    Note that the reflectivity of the 140K data is shifted upwards by 0.6, for clarity.
    The inset highlights the soft-mode response with low pressure data taken in a different set-up down to \icm{40} (see text).}
  	\label{figR}
\end{figure}

We measured the infrared reflectivity of a KDP single crystal in Bruker IFS 113v and 66v/S interferometers. 
The absolute value of the reflectivity was obtained in an ARS Helitran cryostat utilizing an \textit{in situ} gold overfilling technique \cite{Homes1993}. 
We obtained data in the ferroelectric and paraelectric phases with the electric field of light polarized along the ferroelectric axis between 10 and \icm{6\,000}. 

High pressure infrared reflectivity data, up to \GPa{2.7}, were collected at the SMIS beamline, SOLEIL Synchrotron, in a home-made horizontal microscope with custom objectives. 
The microscope is coupled to a Thermo-Fisher iS50 interferometer with synchrotron radiation as its light source; a solid substrate beamsplitter; and a 4K Si:B bolometer to cover the 100 to \icm{600} region. 
We employed a membrane-driven diamond anvil cell (DAC) with a \microns{600} diameter culet, a \microns{60} thick pre-indented stainless steel gasket with a \microns{250} diameter hole filled with polyethylene as hydrostatic medium \cite{Celeste2019}.
The sample was placed in contact with one of the culets and we employed a ruby sphere for the \textit{in situ} pressure measurement. 
The DAC was attached to the cold head of a $\ell$\ce{N2} flow cryostat. 
We utilized the diamond table as a reference. 
The thus-obtained reflectivity in the DAC at the lowest pressure was then corrected to match the measurement taken outside the cell at the same temperature.
We applied this correction function to the other higher pressure measurements.

%
%
The data below \GPa{1} were measured at BNL within a clamped pressure cell using kerosene as the pressure transmitting medium; the entire arrangement sits at the bottom of a compact open-flow cryostat.
This cell employs a wedged diamond window ($\simeq 7^\circ$) with a working diameter of about 1.5~mm, allowing measurements down to $\simeq 20$ -- \icm{30} on a Bruker 66v/S using conventional infrared sources.
The reflectivity unit uses toroidal mirrors with a long focal length.
Rotations allow the reflectivity from the sample in the cell to be separated from the reflectivity obtained from the front of the diamond window, which, similar to the approach used with the DAC, is then used as an optical reference \cite{Kezsmarki2007}.

%
%

Figure~\ref{figR} shows the infrared reflectivity of KDP in the paraelectric and ferroelectric phases. 
We show the data for the full phonon energy range at zero pressure and for a more restricted range at \GPa{2}. 
The thin lines are the data and the thicker lines are fits utilizing $R = \left| (\sqrt{\varepsilon} - 1) / (\sqrt{\varepsilon} + 1) \right|^2$ and the factorized form of the dielectric function \cite{Kurosawa1961}:
%
%
\begin{equation}
    \frac{\varepsilon(\omega)}{\varepsilon_\infty} = \prod_k \frac{\omega_{{\rm LO}_k}^2 - \omega^2 - i \gamma_{{\rm LO}_k} \omega}{\omega_{{\rm TO}_k}^2 - \omega^2 - i \gamma_{{\rm TO}_k} \omega} .
    \label{eqLST}
\end{equation}
$\omega_{\rm TO}$ and $\omega_{\rm LO}$ are the transverse optical (TO) and longitudinal optical (LO) resonance frequencies for each phonon $k$. 
As this model allows for different dampings ($\gamma_{\rm TO}$ and $\gamma_{\rm LO}$) at each of these frequencies, it can better describe anharmonic effects than a Lorentz oscillator model. 
In order to properly fit the data under pressure, we utilized the \GPa{0} spectra below \icm{100} and above \icm{600}. 
We kept the same extension of the data range to perform Kramers-Kronig analysis of the reflectivity.

%
%
\begin{figure}
	\includegraphics[width=8cm]{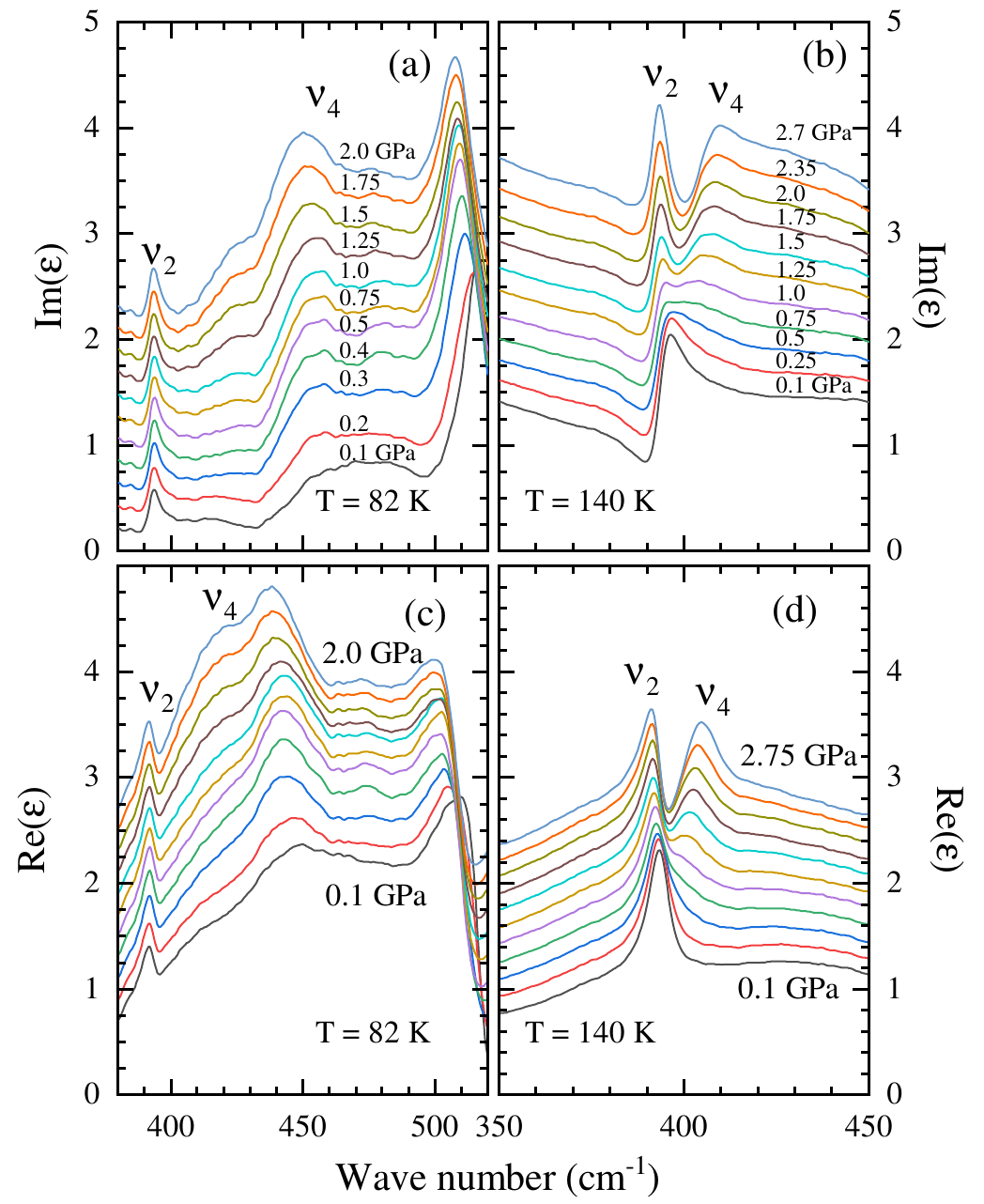}
  	\caption{(color online) Kramers-Kronig obtained dielectric function around the $\nu_2$ and $\nu_4$ vibrational modes of the \ce{PO4} tetrahedron. 
   Panels (a) and (b) show the imaginary part of the dielectric function. 
    Data in panel (a) is taken at \K{82} and in panel (b) at \K{140}. 
    For clarity, the baseline of each curve is increased by 0.2 upon increasing pressure. 
    It is worth noting that in panel (b) $\nu_2$ has a strong asymetric Fano profile at low pressures and becomes a more symmetric conventional phonon at high pressures. 
    Panels (c) and (d) show the corresponding real part of the dielectric function. 
    Here, again, each spectrum is shifted upwards by 0.2 with respect to the previous pressure. }
  	\label{fignu24}
\end{figure}
We are interested in the behavior under pressure of the \ce{PO4} tetrahedra $\nu_4$ mode, situated around \icm{400}. 
Figure~\ref{fignu24} shows the Kramers-Kronig obtained imaginary part of the dielectric function in the $\nu_4$ vicinity as a function of external pressure for the ferroelectric and paraelectric phases, at \K{82} [panel (a)] and \K{140} [panel (b)]. 
These figures also show the $\nu_2$ mode, which is not related to the ferroelectric polarization but that will help us to understand the behaviour of the $\nu_4$ mode. 

At both temperatures, the $\nu_4$ mode gets narrower upon increasing pressure. 
At \K{82} and \GPa{0.1} one can barely discern a broad peak around \icm{450}, which becomes increasingly sharper when moving towards \GPa{2}. 
In the paraelectric phase one observes a similar effect. 
The $\nu_4$ mode is hardly visible at \GPa{0.1} and, upon increasing pressure, it develops a sharp step creating a prominent dip around \icm{400}. 
At this temperature, the $\nu_2$ mode also shows a particular behavior. At \GPa{0.1} it has a very asymmetric line shape and at \GPa{2} it recovers a regular phonon symmetric profile. 
As we will discuss further, this can be understood as a Fano coupling to the highly-damped $\nu_4$ mode broad background, that disappears when the $\nu_4$ vibration becomes less damped.

Panels (c) and (d) show the real part of the dielectric function corresponding to the data in panels (a) and (b).
At \K{82} the $\nu_4$ mode develops a sharper profile. 
At \K{140} we can only devise a single excitation at low pressures, whereas two clear peaks for the $\nu_2$ and $\nu_4$ modes appear at high pressures. 

%
%

In order to quantify our results, let us take a look at the fitting parameters at \K{140}, where the effect is more spectacular. 
For this, besides an analysis utilizing the factorized form of the dielectric function from Eq.~(\ref{eqLST}), we also looked at the aforementioned asymmetric lineshapes.
We looked at the imaginary part of the dielectric susceptibility $\chi_F^{\prime\prime}$ with a Fano profile \cite{Davis1977}:
%
%
\begin{equation}
    \Delta \chi_F^{\prime\prime} = R \left[ \frac{(\omega_F - \omega - \gamma_F  q)^2}{(\omega_F - \omega)^2 + \gamma_F^2} - 1 \right] .
    \label{eqFano}
\end{equation}
Here, $\omega_F$ is the renormalized phonon frequency, $\gamma_F$ its damping, $R$ a transition rate, and $q$ the Fano-Wigner-Breit parameter. 
As $q^{-2}$ approaches zero, the line shape becomes more symmetric.
The $\Delta$ symbol emphasizes that a Fano oscillator does not exist by itself. 
It must sit on top of a larger background as a differential correction to the dielectric function. 
In our case, we utilized two large Lorentzian oscillators in the background to describe surrounding phonons. 
The Lorentzians, within uncertainties, do not change as a function of pressure. 

\begin{figure}
	\includegraphics[width=8cm]{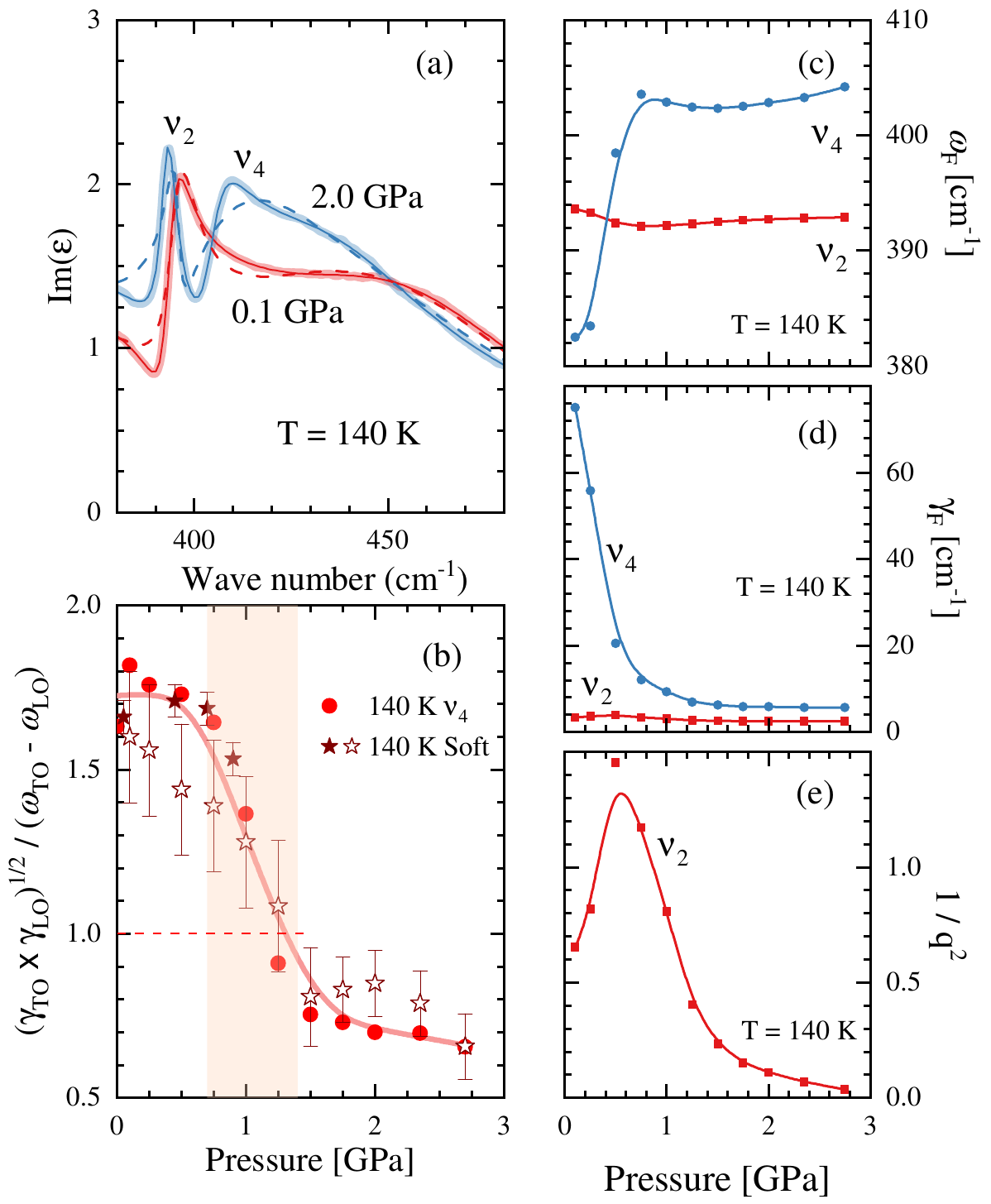}
  	\caption{(color online) (a) Imaginary part of the dielectric function for the $\nu_2$ and $\nu_4$ modes at 0.1 and \GPa{2}.
    The thick lines come from a Kramers-Kronig transformation of the reflectivity. The solid thin lines are fits utilizing Fano 
    profiles [Eq.~(\ref{eqFano})] for these modes.  The dashed lines fits come from the factorized dielectric function 
    [Eq.~(\ref{eqLST})]. 
    (b) Geometric average of TO and LO dampings normalized by the TO--LO frequency splitting, extracted from the factorized 
    form of the dielectric function, for the $\nu_4$ mode (solid circles and solid triangles).  The stars are the same quantity 
    for the soft-mode from data taken at Brookhaven (solid) and Soleil (open).  The soft-mode data was re-scaled by a factor 1.4 
    in order to compare with the $\nu_4$ damping. (c)--(e) Fano parameters for the $\nu_2$ and $\nu_4$ modes.} 
  	\label{figParameters}
\end{figure}

The thicker lines in Fig.~\ref{figParameters}(a) show the Kramers-Kronig obtained imaginary part of the dielectric function at \K{140}. 
For each pressure we show two fits. 
The thin dashed line utilizes Eq.~(\ref{eqLST}). 
It gives an acceptable overall description of the data but fails to describe the sharper line shapes. 
The thin solid lines are from Eq.~(\ref{eqFano}) and follow the data almost perfectly.
%
%

We begin with the imperfect fits coming from the factorized form of the dielectric function. 
It is useful to define the criterion $\sqrt{\gamma_{\rm TO} \gamma_{\rm LO}} > (\omega_{\rm LO} - \omega_{\rm TO}$). 
This is the limit where, in the absence of other phonons, the real part of Eq.~(\ref{eqLST}) becomes positive for all frequencies between $\omega_{\rm TO}$ and $\omega_{\rm LO}$, killing the Reststrahlen phonon scattering.
Figure~\ref{figParameters}~(b) shows that at \K{140} this quantity drops sharply in the pressure region where \citeauthor{Peercy1973} \cite{Peercy1973} observed the underdamping of the Raman soft mode. 
In particular the ratio between the averaged damping and the TO-LO splitting crosses the critical value of 1. 
We also show this normalized damping value for the soft mode. 
As the Brookhaven data stops at \icm{40} and the Soleil data at \icm{100}, we have a large error bar for this quantity. 
Nevertheless, the temperature evolution for the soft-mode damping shows the same trend as the one observed for the $\nu_4$ mode.

Figures~\ref{figParameters}~(c), (d) and (e) show the results of Fano fittings to $\nu_2$ and $\nu_4$ at \K{140}. 
The resonance frequency and linewidth of the $\nu_4$ change remarkably at a pressure lower than \GPa{1}. 
The $\nu_4$ frequency [Fig.~\ref{figParameters}~(c)] hardens significantly at the same pressure that this mode narrows by almost one order of magnitude [Fig.~\ref{figParameters}~(d)].
This highlights the predicted effect of this phonon destabilization by the proton tunneling \cite{Simon1985,Simon1988}, which goes away with pressure.

Looking at the  $\nu_2$ phonon, we observe that both its frequency and linewidth are essentially pressure independent. 
However, $q^{-2}$ [Fig.~\ref{figParameters}~(e)] for this mode changes dramatically around \GPa{0.6}. 
It shows $\nu_2$ going from a very asymmetric to a fully-symmetric line shape. 
As $\nu_2$ and $\nu_4$ are vibrational modes of the \ce{PO4} group, they are naturally coupled when they have close resonance frequencies 
and hence the Fano asymmetric profile. 
Close to ambient pressure, the resonance frequency of $\nu_4$ is smaller than that of $\nu_2$. 
The $\nu_4$ frequency crosses $\nu_2$ at a pressure close to \GPa{0.6}, at the same value where its Fano damping decreases by an order of magnitude. 
This means that $\nu_4$ gets narrow and moves away from $\nu_2$ upon increasing pressure. 
This better separation of the modes is what makes the $\nu_2$ more symmetric. 
Note that $\nu_2$ has no particular role in the ferroelectric transition. 
Here it serves as a probe to the narrowing of $\nu_4$.


%
%

In summary, we measured the infrared reflectivity of \kdp\ under pressure up to \GPa{2.7}. 
We found that the $\nu_4$ vibration mode of the \ce{PO4} tetrahedron is strongly damped at ambient pressure and it becomes underdamped upon increasing pressure. 
Its behavior closely follows the underdamping observed by \citeauthor{Peercy1973} for the KDP soft-mode in Raman scattering, indicating a coupling between these two modes. 
The $\nu_4$ mode, which has an electric dipole and involves only heavier P and O ions, is strongly destabilized by the presence of the nearby protons. 
It is the missing piece of KDP soft-mode as it represents the excitation responsible for coupling the hydrogen network to the lattice modes.
In particular it explains how the in-plane proton dynamics affects the out-of-plane phonon at the origin of the electric polarization.

\acknowledgments

Measurements at Soleil Synchrotron were carried out at the SMIS beamline under proposal 20220159.
A.A. acknowledges funding from the  Swiss National Science Foundation through project PP00P2\_202661.
Work at Brookhaven National Laboratory was supported by the Office of Science, U.S.
Department of Energy under Contract No. DE-SC0012704.

%
%

%
%
\input{biblio.bbl}

\end{document}

%% file: biblio.bbl
%